# Explainable AI in Healthcare: to Explain, to Predict, or to Describe?

**Alex Carriero**
University Medical Center Utrecht

**Anne de Hond**
University Medical Center Utrecht

**Bram Cappers**
Eindhoven University of Technology

**Fernando Paulovich**
Eindhoven University of Technology

**Sanne Abeln**
Utrecht University

**Karel GM Moons**
University Medical Center Utrecht

**Maarten van Smeden**
University Medical Center Utrecht

August 7, 2025

## Abstract

Explainable Artificial Intelligence (AI) methods are designed to provide information about how AI-based models make predictions. In healthcare, there is a widespread expectation that these methods will provide relevant and accurate information about a model's inner-workings to different stakeholders (ranging from patients and healthcare providers to AI and medical guideline developers). This is a challenging endeavor since what qualifies as relevant information may differ greatly depending on the stakeholder. For many stakeholders, relevant explanations are causal in nature, yet, explainable AI methods are often not able to deliver this information. Using the Describe-Predict-Explain framework, we argue that Explainable AI methods are good descriptive tools, as they may help to describe *how* a model works but are limited in their ability to explain *why* a model works in terms of true underlying biological mechanisms and cause-and-effect relations. This limits the suitability of explainable AI methods to provide actionable advice to patients or to judge the face validity of AI-based models.

# 1 Introduction

In healthcare, there is a long tradition of developing prediction models to estimate the probability that a given health outcome is present (diagnosis) or will occur in the future (prognosis) on an individual patient basis[1]. This information can be helpful to both healthcare providers and patients. For example, a cardiologist might use SCORE-2 to estimate the risk that a patient will develop cardiovascular disease in the next ten years[2]. Proponents claim that the recent developments in AI will make such risk estimation become even more accurate and tailored to the specific patient as more diverse data sources are used (e.g., images, text) and as more complex models are considered[3]. Meanwhile criticism of more these more complex models is often related to their black-box nature, especially when used in sensitive settings like healthcare[4–8].

A variety of explainable AI techniques have been suggested to help circumvent the black-box nature of AI, making the inner-workings of AI-based prediction models more transparent. Examples of explainable AI methods include the commonly reported Local Interpretable Model-Agnostic Explanations (LIME), SHapley Additive exPlanations (SHAP) and Gradient Class Activation Mapping (GradCAM)[5,6,9–11]. The aim of these methods is to provide understandable and relevant information about an AI-based prediction model's input and functionality. However, the degree to which the information from more transparent AI is useful and relevant may depend on the stakeholder receiving the information, and acting upon it[12]. Moreover, as we will argue in this paper, the level of transparency that can be reached by explainable AI methods may not align with what important stakeholders, such as healthcare professionals and patients, want or expect from these methods.

In particular, stakeholders may seek information that is causal in nature, since evidence-based (causal) information is typically necessary for medical decision-making. Yet, current explainable AI methodologies are ill-equipped to deliver such causal guarantees when applied to explain the predictions of correlation-based prediction models[13]. For example, if a cardiologist communicates to a patient that their risk of cardiovascular disease is high based on a prediction from the SCORE-2 model, it is highly desirable for a patient to understand *why* they are at high risk, and what they might be able to *do* to improve their prognosis; these are inherently causal questions. With explainable AI methods, it is technically feasible to determine what patient characteristics led to the prediction and what input values could change in order for the model to give a lower risk estimate. However, guidance based on explainable AI output often does not come with causal guarantees, meaning that it may not align with guidance that is based on true cause-and-effect relationships. In other words, the guidance may be ineffective or provide potentially harmful recommendations. Thus, information derived from explainable AI methods may not meet the standards necessary for use in medical decision-making.

In this paper, we use the Describe-Predict-Explain framework introduced by Shmueli[14] to highlight the type of information that explainable AI generally provides. We argue that within the framework, explainable AI generally provides descriptions, not explanations. Using an illustrative example we demonstrate the risks of misinterpreting the results of explainable AI methods for health(care) decision making.

# 2 Framework: Describe, Predict or, Explain

The distinction between description, prediction and explanation is foundational in the fields of statistics and data science[14,15], representing three distinct motivations for why one might develop a statistical model (e.g., a regression or machine-learning based model).

If the objective is *description*[14], the aim is to describe patterns in available patient data. For example, medical researchers may compute summary statistics, correlations or develop a (multi-variable) model which relates patient characteristics to a health outcome (e.g., disease, side effect, treatment effect, etc.). In the latter case, the interest lies in studying how the model combines the features in order to compute the predicted outcome. The expression the model uses to relate patient characteristics to a health outcome is seen as a description of the data since it captures relations between the features and the outcome observed in the model development data. For example, given data from multiple prospective cohorts following breast cancer diagnosis, descriptive information may include a summary of what characteristics are more common in women with breast cancer vs. those without breast cancer[16].

If the objective is *prediction*[14], the aim in healthcare is often to develop a prediction model that gives accurate risk predictions for a diagnostic or prognostic outcome in individual patients[1,17]. In this context, it is helpful to think about clinical prediction models as tools for use by healthcare professionals and patients. These tools should be easy to use (e.g., involve readily available information) and function effectively in the setting where they will be applied (e.g., provide accurate, reliable and clinically useful predictions). The quality of the predictions for future patients is the primary concern. Examples of such prediction models can be found in almost every field of medicine, from disease prevention, to care, and to cure. For example, a prediction model might be used to estimate the risk of femur fracture from a patient's radiograph (diagnostic model)[18] or to estimate a patient's 10-year risk of fatal cardiovascular disease based on their age, smoking status, systolic blood pressure, total- and HDL-cholesterol (prognostic model)[2].

If the objective is *explanation*, the aim is to gain an understanding of the world around us[17]. In healthcare this often equates to understanding the direct, causal effect of a patient characteristic, exposure or an administered treatment on patient health outcomes. Intervention studies (i.e., randomized experiments) are generally perceived as the gold standard for causal evidence in medicine. When no randomization is possible or ethical, causal effects may be studied from non-randomized studies using causal inference methodologies. In this domain there is a specific focus on answering *"What if?"* questions with respect to an actionable intervention[19]. For example, when estimating a patient's 10-year risk of fatal cardiovascular disease, one may be interested in studying the consequence of an actionable change (e.g., an intervention) on the patient's outcome: *what if* a given patient stops smoking?

## 3 To describe or explain AI-based prediction models

AI-based prediction models, unsurprisingly, align most clearly with Shmueli's motivational domain of prediction. However, stakeholders (i.e., model developers and model users) may require information that is not clearly aligned with the domain of prediction. Instead, model developers and users often seek information about *how* a given prediction model functions (i.e., how did the input information influence the a patient's risk estimate). To which domain does this information belong? I.e., do explainable AI methods provide, descriptions or explanations? We argue that the results of explainable AI methods align closely with Shmueli's domain of description.

The domain of explanation is different from description as the aim is to understand cause-and-effect relationships. While AI-based prediction models that function effectively at their prescribed task may provide the illusion that they have an internal understanding of the world, these models have no ability to distinguish causes and effects, and are most often trained to learn statistical associations apparent in observational data. These prediction models are often developed by minimizing the difference between the model's predictions (e.g., estimating risk of cancer in 5 years) and observed outcomes (e.g., observation of whether or not a cancer diagnosis was given in 5 years) across all individuals in a data set chosen for model development. This optimization objective can be accomplished without regard for causal knowledge[20]. Thus, if the prediction model is trained to learn statistical associations in the model development data, any corresponding model explanations will then align well with Shmueli's domain of description, not explanation.

Consider the earlier example from the domain of description: what characteristics are more common in women with breast cancer vs. without breast cancer given data from many prospective cohorts following a breast cancer diagnosis? For a prediction model developed using a database of observations from these cohorts, the same questions can be reformulated as: what patient characteristics contribute to the prediction of breast cancer? In this case, the prediction model makes predictions based on patterns apparent in the model development data. Thus, information about how the model works is, by extension, a description of the model development data. As we illustrate next, these descriptions or model "explanations" may or may not align with true underlying cause-and-effect relationships, depending on what variables were measured and included in the model.

Even though causal knowledge is not required to develop a prediction model, it is possible, and often very helpful, for model developers to incorporate causal knowledge into a prediction model. This might include using domain knowledge to determine what information should be included in the model (i.e., what predictors should be measured) or to determine how best to incorporate existing treatment strategies in the model[17,21]. Prediction models may also be used to directly answer "What if?" questions with respect to a specific intervention (e.g., prediction under intervention models)[22,23]. Yet, including causal information during model development or using a prediction model to estimate a *single* causal effect does not necessarily improve our ability to interpret corresponding model explanations for *all* included variables as causal. This is because the assumptions necessary to estimate a causal effect are typically met for at most one variable (e.g., treatment), not all variables included in the model.

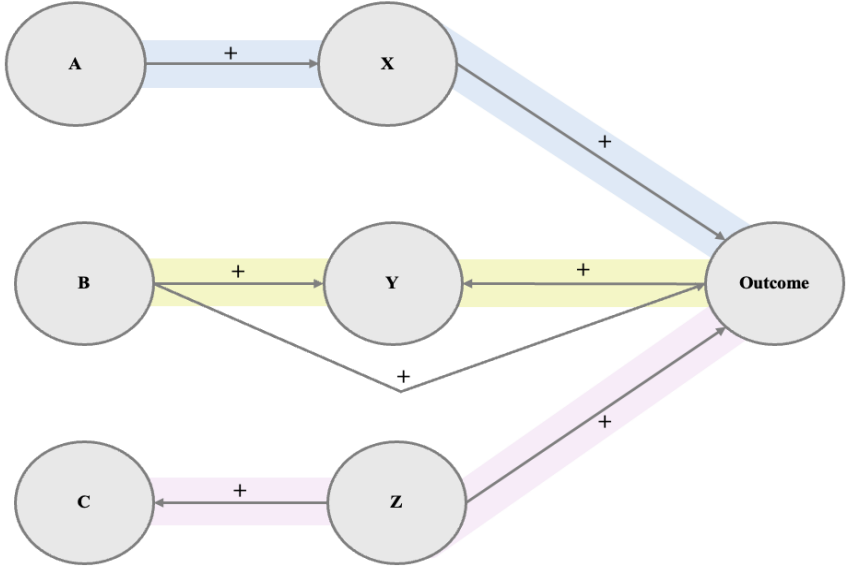

Figure 1: Data-Generating Mechanism. Arrows with + indicate a positive causal effect.

## 4 Illustration

The phrase "*correlation does not equal causation*" is widely remembered, yet, the consequences of the this statement are not necessarily intuitive. Using an illustrative example we highlight how different multi-variable cause-and-effect relationships can result in data sets where the correlations present do not match the data-generating (causal) mechanism (e.g., correlation with opposite sign to causal effect), yet still result in a model with adequate predictive performance. In this illustration, we created a simple data-generating mechanism that captures three causal structures (Figure 1). We generated data accordingly, developed a black-box prediction model and validated the model's performance. Subsequently, we studied the inner-workings of the prediction model using a common post-hoc explainable AI method. The code for this illustration is freely available on GitHub, https://github.com/alexcarriero/illustrative_example.

In Figure 1, we see that in our data-generating mechanism the Outcome has three direct causes X, B, and Z, represented by arrows terminating at the Outcome. First, consider the path highlighted in blue. Here we see that variable A causes variable X which in turn causes the Outcome. This might represent a dynamic such as: smoking (A) causing increased hypertension (X) which in turn causes increased risk of kidney disease (Outcome). Next, consider the path highlighted in yellow. In this case, we see that both variable B and the Outcome cause variable Y. This might represent a dynamic such as: age (B) and kidney disease (Outcome) both causing increased risk of hospitalization (Y). Finally, let's consider the pathway in pink. Here variable Z is a common cause of both variable C and the Outcome. This might represent a dynamic such as: diabetes (Z) causing kidney disease (Outcome) and also causing a patient to be prescribed insulin (C). In summary, our data-generating process includes three pathways, each representing distinct types of causal relations among two variables and the outcome (i.e., intermediate, collider, and confounder, respectively).

Consider the following hypothetical AI model to predict a diagnosis of kidney disease, within a group of consecutive suspected patients admitted to a hospital. We are able to measure the candidate predictors (input features): patient smoking status (A), hypertension (X), age (B) and if the patient has been prescribed insulin (C). Given we are studying hospitalized patients (a subgroup of our original population), information about variable Y is now implicitly included in the prediction model we develop (i.e., since we have restricted our analysis to patients in-hospital).

A prediction model for the diagnosis of kidney disease with the candidate predictors was developed on a large simulated data set of 20,000 observations. We used XGBoost to develop the prediction model[24] and assessed model performance using a large (n = 100,000) independent validation set generated using the same data-generating mechanism; this model had an AUROC of 0.80 and was well-calibrated. Shapley value explanations were subsequently generated for all individuals in the model development data[9] (Figure 2). We used TreeSHAP[25] to compute the SHAP values and the reference value was derived from a random sample of 5000 individuals from the development data. For a brief introduction to SHAP see TextBox 2. To see the complete model development and validation code and results please see our codebook.

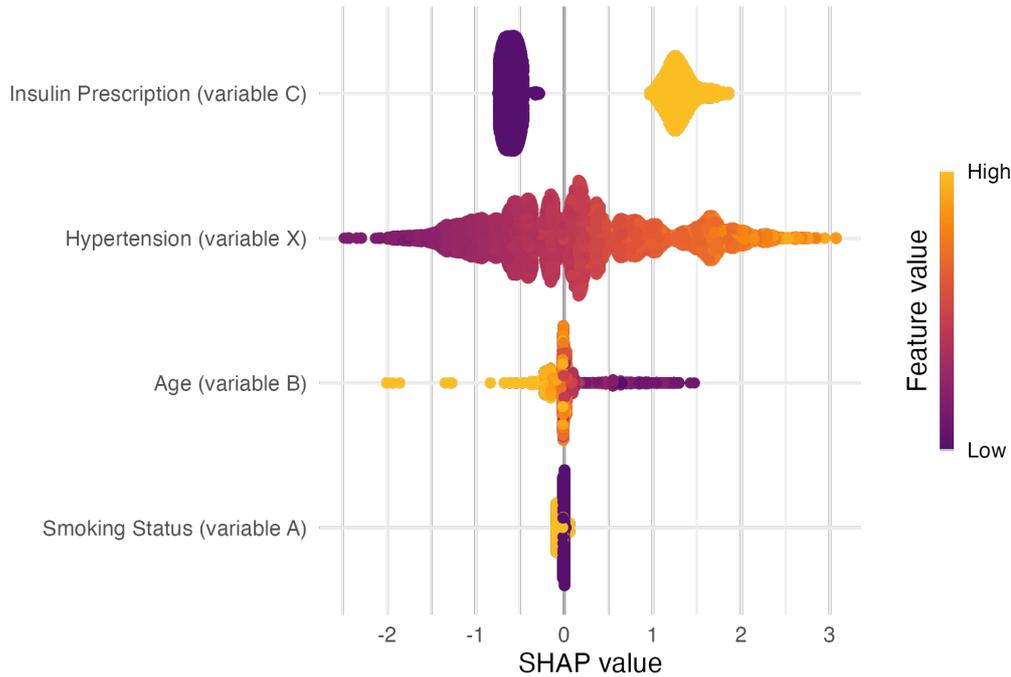

Figure 2: SHAP bee-swarm plot for our illustrative example. Every dot represents a SHAP value for given feature of an individual in the model development data. On the horizontal axis, SHAP values that deviate from 0.00 indicate a positive or negative associations with the model's predictions. On the vertical axis, the features are ranked from top to bottom based on their importance to the model's predictions. Purple dots correspond to low feature values and yellow dots to high feature values. For example, individuals who were prescribed insulin (yellow dots in the insulin prescription row) have positive SHAP values, indicating that they had higher predicted risks than those who were not prescribed insulin.

In Figure 2, we see the output of the chosen explainable AI method (a SHAP bee-swarm plot). This plot provides insight into the model's overall functionality. Since we know the true causal data-generating mechanism for our data-set, we can compare these insights with our expectations based on the true data-generating mechanism.

### 4.1 Intermediate in the causal chain

Based on the data-generating mechanism, we know that smoking status (A) has a causal relationship with the occurrence of kidney disease (Outcome). Yet, smoking status is identified as the predictor with the least effect on the prediction. From Figure 2, we see that hypertension (X) contributes much more to the model's predictions than smoking (A), which is the original cause and hypertension is the intermediate variable (see Figure 1). Of course, it is misleading to conclude that smoking is not a cause of kidney disease based on the model explanation. The correct interpretation here is that, once information about hypertension is present, the information about smoking is less or even no longer helpful in the prediction model, since the information contained by hypertension and smoking is largely interchangeable for making the prediction of kidney disease.

## 4.2 Collider

From Figure 2, we observe another surprising result: age (B) has a negative relation with the risk of kidney disease (i.e., higher age means lower risk of kidney disease). In our causal structure (Figure 1), we specified that this relation is positive (i.e., higher age means higher risk of kidney disease). This illustrates a phenomenon called "collider bias". In this case, by restricting individuals in the development data based on variable Y (e.g., including only hospitalized patients) we introduced a spurious correlation between age and kidney disease. This correlation is spurious in the sense that it is not seen in the general population (where the opposite is actually true). Yet, this spurious correlation is completely representative of the development population (e.g., younger people who find themselves in the hospital might have a higher incidence of kidney disease than older people who may be in the hospital for many other reasons). Thus, the negative relation is an accurate description of the development data, but not of the causal mechanism of age and kidney disease. Therefore, while the correlation for this variable likely does not generalize outside of hospitalized patients, it could be worth including in the model under the premise that the model will be deployed in similar hospital settings. In other words, correlations that show a direction contraindicative to the expected direction are not necessarily a clue that the prediction model is not functioning well, they may be the result of collider bias.

## 4.3 Confounder

Finally, consider the predictor insulin prescription. While we know that insulin use (C) has no causal relationship with kidney disease (Outcome), it has the strongest association with the outcome among the candidate predictors. The correlation learned by the model is the result of the (unmeasured) confounder diabetes, which was not considered in this example as a candidate predictor. In other words, a strong relation between insulin and the Outcome is introduced as a result of not recording diabetes diagnosis in the development population and subsequently, not including it as a predictor in the prediction model. Prescription of insulin does not cause kidney disease (Figure 1), if we had measured and included information regarding a patient's diabetes diagnosis, the prescription of insulin would likely cease to be an important predictor of kidney disease. This illustrates that using model explanations as the basis for actionable medical decisions can also be dangerous (e.g., ceasing to prescribe insulin would certainly not prevent people from getting kidney disease and recommending that patients stop taking insulin would likely worsen their health).

Through the above illustration, we hope to have compelled the reader that the patterns apparent in the development data of prediction models are largely affected by which variables were measured and how those variables relate to each other and the outcome in the underlying causal pathways. Prediction models learn to predict an outcome based on patterns in the development population and in turn, model explanations aim to recover the patterns. While the model explanations in this example do not *explain* the data-generating process, they do *describe* the patterns apparent in the development data.

## 5 Are all prediction model explanations "just" descriptions?

Within the domain of explainable AI there are two common strategies. The first is to create a prediction model that is simple enough for a human to understand e.g., prediction based on a simple decision tree or sum score (e.g., simple rules for predicting ICU congestion[26]). Alternatively, if prediction models are more complex (i.e., the modeled relations between the input features and predictions are not clear), then one may apply post-hoc methodologies (e.g., SHAP and LIME[9,10]) to elucidate information about how the model makes predictions. Note however that the distinction between "simple" and "complex" models is subjective (see TextBox).

The post-hoc explainable AI methods typically prioritize human understanding over a faithful (completely correct) description of how the model computes predictions. Since humans are not typically adept at comprehending multi-dimensional relationships, information about important characteristics of the complex model (e.g., strong interaction effects, non-linearities) can be lost in favor of an explanation which is more easily understood[7,20,27,28]. This is disconcerting as the complex patterns that warrant the used of a complex model over a simpler model may remain completely hidden. Furthermore, limitations of the post-hoc methods can result in substantial added uncertainty about the degree of agreement between the information extracted via the post-hoc methods of the true inner-workings of a more complex prediction model[7,28,29].

Although a post-hoc explainability method was used in our illustration, we emphasize that our conclusions hold more generally, even for models that are fully transparent by design. As long as a models relies on learned statistical associations from a data set to make predictions, then the aim of explainable AI techniques, in general, is to uncover the patterns learned by the model from this data set. Regardless if the patterns are directly apparent, as they are with simple models, or if they are elucidated by post-hoc techniques (albeit with some added uncertainty), the results do not offer support of causal effects unless accompanied by the necessary assumptions[19]. For instance, for more transparent models like regression models, this realization is sometimes called the Table 2 Fallacy[30]; it is not advisable to interpret all model coefficients in a multivariable regression model as causal effect estimates as typically the assumptions necessary to estimate a causal effect are met for at most one variable (e.g., treatment), not all variables included in the model.

## 6 Discussion

In conclusion, we argue that explainable AI methods align best with Shmueli's domain of description, not explanation, when used with correlation-based prediction models; explainable AI methods do not distill causal information out of a non-causal model. Descriptions summarize patterns apparent in a data set and such patterns are largely affected by which variables were measured and how those variables relate to each other and the outcome in terms of cause-and-effect relationships. Misinterpreting model explanations as causal explanations can be misleading and lead to misinformed decision-making. Therefore, for stakeholders with causal interests (e.g., model users seeking an explanation for a prediction that is congruent with true underlying biological mechanisms and cause-and-effect relations) descriptive information is not suitable.

In our illustration, it was easy to see that the relationships learned by the prediction model, and subsequently presented to us via the model explanations did not match the data-generating mechanism. Given our understanding of the data-generating mechanism, we could provide a rationale for why certain correlations were learned (or not learned) by the model, and could examine which correlations matched the causal effects and which did not. In practice, when there may not be 6 but rather, hundreds of predictor variables available, the pursuit of understanding which relationships we can expect to align with the data-generating mechanism (and which we cannot) becomes exponentially more difficult. This is further complicated by an incomplete understanding of the (causal) data-generating mechanisms in many clinical applications. While many recent methods aim to provide causal explanations[31–34], these methods necessarily assume knowledge of a causal graph (a diagram representing the causal relationships between all relevant variables and the outcome) which is rarely available for multi-variable clinical prediction models in practice[35].

In this paper, we sought to highlight that good prediction models can have surprising correlations (opposite from causal direction) and consequently, model explanations can be poor resources for judging face-validity of a model or for giving actionable advice to patients. However, there are many other reasons to opt for simple transparent models and/or to generate model explanations (e.g., ease of implementation, transparency in general, hypothesis generation, exploratory analyses, auditing). Depending on the specific context of a prediction model, we encourage discussion about whether or not explainable AI is necessary, how explainability should be achieved (simple and interpretable models vs. post-hoc methods) and ethical dimensions about the use of machine learning in healthcare.

In summary, explainable AI techniques have the aim of lessening the risks associated with black-box modeling by providing more or fully transparent AI-based models, yet, when explainable AI is misused or misinterpreted new risks are introduced. Hence, a clear understanding of the boundaries of these methods is necessary for the safe deployment of prediction models in healthcare.

**Author Contributions**:

Concept and Design: AC, MvS, AdH, BC

Drafting Manuscript: AC, MvS

Editing / Critical revision of manuscript for intellectual content: all authors.

**Acknowledgments**:

The authors kindly acknowledge Florian van Leeuwen and Lotta Meijerink for helpful feedback and discussions regarding our illustrative example.

**False Dichotomy (TextBox 1)**

The division of models into "glass box" (or "inherently interpretable") and "black box" models represents a false dichotomy. The line between "glass box" and "black box" models is necessarily subjective. It cannot be drawn on the basis of which algorithm is used to develop the prediction model, as even decision trees and logistic regression models can easily also become too complicated to understand (e.g., multi-feature interactions, hundreds of features), while for instance neural networks may be constrained such that they can be easily understood. Rather, the distinction is dependent on the interpreter, and is subjective as it may vary among stakeholders. While there are certainly models for which no human will have intuition, and likewise, very simple decision trees which all stakeholders may understand, between the extremes there exist models which may be seen as interpretable to some stakeholders while not to others.

**Introduction to SHAP (TextBox 2)**

Shapley value explanations are model explanations that are generated on an individual basis. The explanations are presented as a set of feature contributions i.e., each input feature in the model is given a SHAP value indicating how the observed value for that feature affected a given individual's prediction. Shapley value explanations in our example may be interpreted as an explanation for why a prediction differs from the global average prediction (where the global average prediction is estimated using a reference population e.g., a random sample from the development population). The SHAP values for each individual have a nice additive property: the difference between an individual's prediction and the global average prediction is distributed perfectly across the feature contributions (SHAP values sum to the difference between an individual's prediction and the global average prediction). When SHAP explanations are generated for many individuals, they may be visualized together to help elucidate information about the model's overall functionality. A beeswarm plot (Figure 2), presents the SHAP values for many individuals organized by each feature in the model. In this plot the features are organized from top to bottom based on their importance to the model's predictions.